\documentclass[aps,prl,superscriptaddress,twocolumn]{revtex4-2} 

\usepackage{amssymb} 
\usepackage{amsmath}
\usepackage{epsfig}
\usepackage{graphics, graphicx}
\usepackage{bbold}
\usepackage{psfrag}
\usepackage{mathcomp}
\usepackage{subfigure}
\usepackage{verbatim}
\usepackage{color}
\usepackage[colorlinks=true,linkcolor=blue,citecolor=blue,urlcolor=blue]{hyperref}

\begin{document}
 
\title{Efimov physics in the complex plane}

\author{Mingyuan Sun}
\affiliation{State Key Lab of Information Photonics and Optical Communications, Beijing University of Posts and Telecommunications, Beijing 100876, China}
\affiliation{School of Science, Beijing University of Posts and Telecommunications, Beijing 100876, China}

\author{Chang Liu}
\affiliation{Institute for Advanced Study, Tsinghua University, Beijing 100084, China}
\author{Zhe-Yu Shi}
\email{zyshi@lps.ecnu.edu.cn}
\affiliation{State Key Laboratory of Precision Spectroscopy, East China Normal University, Shanghai 200062, China}
\date{\today}

\begin{abstract}

Efimov effect is characterized by an infinite number of three-body bound states following a universal geometric scaling law at two-body resonances. In this paper, we investigate the influence of two-body loss which can be described by a complex scattering length $a_c$ on these states. Interestingly, because of the complexity of the scattering length $a_c$, the trimer energy is no longer constrained on the negative real axis, and it is allowed to have a nonvanishing imaginary part and a real part which may exceed the three-body or the atom-dimer scattering threshold. Indeed, by taking the $^{133}$Cs-$^{133}$Cs-$^6$Li system as a concrete example, we calculate the trimer energies by solving the generalized Skorniakov-Ter-Martirosian equation and find such three-body bound states with energies that have positive real parts and obey a generalized geometric scaling law. Remarkably, we also find that in some regions these three-body bound states have longer lifetimes compared with the corresponding two-body bound states. The lifetimes for these trimer states can even tend to infinity. Our work paves the way for the future exploration of few-body bound states in the complex plane.   

\end{abstract}

\maketitle

Three identical bosons can form an infinite series of bound states nearby a two-body resonance, which satisfy a universal scaling law and display a discrete scaling symmetry~\cite{Efimov}. It is called the Efimov effect and has been generalized to various three-particle systems with different mass ratios and statistics~\cite{Braaten,Greene,Naidon}. Efimov physics has been observed in a number of cold atom experiments~\cite{Efimov_Exp0,Efimov_Exp1,Efimov_Exp1bu,Efimov_Exp3,Efimov_Exp4,Efimov_Exp5,
Efimov_Exp9,Efimov_Exp10,Efimov_Exp6,Efimov_Exp7,Efimov_Exp8,Efimov_Exp11,rf_1,rf_2,
scaling_1,scaling_2,scaling_3} as well as in gaseous helium~\cite{Helium}. Conventionally, the studies on the few-body physics are restricted in dissipationless closed systems. However, dissipation is ubiquitous in cold atoms. On the one hand, dissipation naturally leads to the decoherence or decay of quantum states. On the other hand, it has been shown that dissipation can also be used as a tool to engineer exotic new physics in open systems~\cite{Durr,topo1,topo2,topo3,topo4,topo5,topo6,topo7,Das,Kawakami,Cui1,Gerbier,Zhai,
Ueda1,Ueda,Iskin,Cui2}. For instance, it can induce correlation in one dimensional molecular gases~\cite{Durr}, a reentrant superfluid transition~\cite{Kawakami,Iskin}, and distinct topological phases with complex energy spectra, which have no Hermitian counterpart in closed systems~\cite{topo1,topo2,topo3,topo4,topo5,topo6,topo7,Ueda}. Moreover, it can also be adopted as a probe to detect the equilibrium property of a Hermitian system~\cite{Gerbier,Zhai}. Thus, naturally one might wonder how dissipation affects few-body physics such as Efimov effect in a three-body system. 


In this work, we shed light on this problem by studying the effect of short-range two-body loss on the Efimov bound state. Two-body loss naturally occurs in ultracold atom systems as the two-atom collision process in the atomic gases is generically inelastic. It is shown that such short-range two-body loss can be described by a complex scattering length $a_c$~\cite{Chin,CLS}. As an analog of the conventional real $s$-wave scattering length, the complex scattering length characterizes the inelastic scattering process between two atoms and requires the two-body wave function having asymptotic behavior $\psi(\mathbf{r})\simeq r^{-1}-a_c^{-1}$ for $r\rightarrow0$. In particular, the value $a_c^{-1}$ can be experimentally tuned across the entire upper half complex plane in ultracold atomic gases by the (optical) Feshbach resonance which couples two ground state atoms to an excited two-atom bound state with finite lifetime~\cite{Chin,Cui2}. This allows us to go beyond the conventional real scattering length constraint in the few-body calculations and explore the Efimov physics in the complex $a_c^{-1}$ plane.

In the following, we focus on a three-body system consisting of two heavy identical bosons and one light particle (boson or fermion). We assume there exist short-range interactions and two-body losses between the heavy boson and the third particle which can be characterized by a complex scattering length $a_c$. And the interaction and loss between heavy bosons is set to be zero for simplicity.  One advantage of this system is that the scaling factor becomes smaller when the mass ratio between the heavy boson (with mass $m_b$) and the light particle (with mass $m_0$) $\eta=m_b/m_0$ increases. This facilitates both the experimental observation and theoretical calculation of the geometric scaling law~\cite{scaling_1,scaling_2,scaling_3}. For example, in $^{133}$Cs-$^{133}$Cs-$^6$Li system, the scaling factor is $e^{2\pi/s_0}\simeq23.7$, compared to 515.0 for three identical bosons. We calculated the energy spectra of three-body bound states in such system by solving a generalized Skorniakov–Ter-Martirosian (STM) equation with two-body losses~\cite{STM, Braaten}. 

\begin{figure}[t] 
    \centering
    \includegraphics[width=0.45\textwidth]{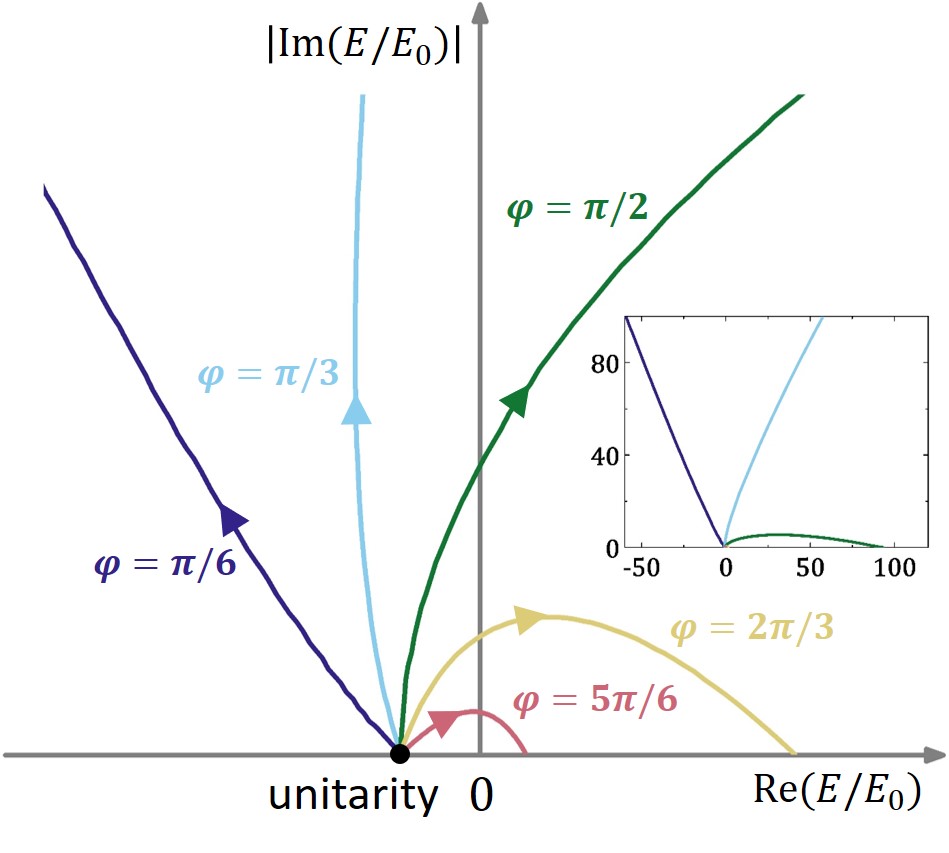}
    \caption{ Energy trajectories of Efimov trimers in the complex plane. For each trajectory, we fix the argument of complex scattering length $\phi=\text{arg}(a_c^{-1})$ and gradually increase its norm $|a_c^{-1}|$ from zero (unitarity) to a positive number. We only show the trimer states that connect to the third lowest Efimov trimers at unitarity. All the trajectories start from the same unitarity point with $E\simeq-0.35E_0$ ($E_0=\Lambda^2/(400m_b)$ is the energy unit), while they vanish at a three-body scattering state ($E\in\mathbb{R}^+$) for $\phi \geq \pi/2$, and at an atom-dimer scattering state ($E\in\{E_d+\Delta E:\Delta E\in\mathbb{R}^+\}$) for $\phi < \pi/2$. We note that the latter scenario is not visible because of its large scale. This will be demonstrated in detail in Fig.~\ref{fig3}. Inset is a zoom-out of the same plot. }
     \label{fig1}
\end{figure}

Our main result is summarized in Fig.~\ref{fig1}. Owing to the two-body losses, the energies of the Efimov trimers are analytically continued from the negative real axis to the complex energy plane. Thus, to display the complex trimer energies as functions of complex $a_c$, we fix the argument $\phi\equiv\text{arg}(a_c^{-1})$ and gradually increase its norm $|a_c^{-1}|$ from zero to a positive number and plot the trajectories of the energies on the complex plane. 
First of all, we find that at $|a_c^{-1}|=0$ or equivalently $|a_c|=+\infty$, the argument $\phi$ does not affect the trimer energies. That is to say, the system only has one resonance point even in the presence of two-body losses, i.e. $a_c^{-1}=0$~\cite{footnote0}. Therefore, all the trajectories start from the same point on the negative real axis which represents conventional Efimov trimer energy at resonance, and continuously extends to the complex plane for different arguments $\phi$. Two different scenarios may occur for the trajectories in the parameter regime $\phi \in [0, \pi]$. For $\phi \geq \pi/2$, it ends at a three-body scattering state with a positive energy $E\in\mathbb{R}^+$, and results in an approximate semi-circle shape in the complex energy plane. One distinct feature from the conventional Efimov effect is that the energy trajectories do not end at the three-body scattering threshold, i.e., $E=0$, but are able to go beyond with Re$(E)>0$ in the presence of dissipation. As $\phi\rightarrow \pi$, the trajectory recovers the Efimov result on the BCS side. On the other hand, for $\phi<\pi/2$, there exists a two-body bound state with complex energy $E_d$. This energy defines the complex atom-dimer scattering threshold, for no three-body bound state exists on the horizontal ray (atom-dimer continuum) defined by $\{E\in\mathcal{C} : {\rm Re} (E)\geq{\rm Re} (E_d) \ \& \ {\rm Im} (E)={\rm Im} (E_d) \}$. The energy trajectories in this region extend beyond the atom-dimer scattering threshold with Re$(E) >$ Re$(E_d)$ and disappears into a state at the atom-dimer continuum. As $\phi\rightarrow 0$, it again recovers the conventional result for Efimov states on the BEC side.

\textit{Model.} Generally speaking, the dynamics of an open system is governed by the Lindblad equation. For a three-particle system, the dynamics is greatly simplified and the Lindblad equation for the three-particle density matrix is fully described by a non-Hermitian Hamiltonian~\cite{CLS}
\begin{equation}
\mathcal{H}=\frac{{\bf p}_{0}^2}{2m_0}+\sum\limits_{j=1,2}\frac{{\bf p}_j^2}{2m_\text{b}}+\sum\limits_{j=1,2}g_c\delta(\mathbf{r}_0-\mathbf{r}_j),
\end{equation}
where ${\bf r}_0$ and ${\bf p}_0$ denote the position and momentum of the light particle, while ${\bf r}_j$ and ${\bf p}_j$ ($j=1,2$) denote the position and momentum of the two heavy bosons. $m_b$ and $m_0$ are respectively the mass of the heavy boson and the light particle. The coupling constant $g_c$ is complex and is related to the complex scattering length $a_c$ by the renormalization relation
\begin{align}
\frac{1}{g_c}=\frac{m_r}{2\pi a_c}-\frac{1}{V}\sum_\mathbf{k}\frac{m_r}{k^2}\label{r_l}
\end{align}
with $m_r=1/(m_0^{-1}+m_b^{-1})$ being the two-body reduced mass. We note that $\text{Im}(g_c)$ must be negative to ensure the positive definiteness of the density matrix, which means $\text{Im}(a_c)\leq0$ or equivalently, the inverse scattering length $a_c^{-1}$ is restricted in the upper half complex plane.

With the renormalization relation, one can calculate two-body scattering $t$-matrix with energy $E$,
\begin{equation}
t_2(E)=\frac{2\pi}{m_r} \frac{1}{a_c^{-1}-\sqrt{-2m_r E}}
 \label{t2}
\end{equation} 
Note that the negative real axis is taken as the branch cut of $\sqrt{\cdot}$.  When Re$(a_c^{-1}) \geq 0$, there is a pole at $E_d=-1/(2m_r a_c^2)$, representing the system can support a two-body bound state with energy $E_d$~\cite{footnote0.5}.

Based on Eq.~\ref{t2}, one can further derive the $s$-wave atom-dimer scattering matrix $ t_3(p,k;E)$ with the relative incoming and outgoing momenta $p$, $k$, and energy $E$, which can be written as
\begin{widetext}
\begin{equation}
  t_3(p,k;E)=\frac{m_0}{2pk}{\rm ln}\frac{E-\frac{p^2+k^2}{2m_r}+\frac{pk}{m_0}}{E-\frac{p^2+k^2}{2m_r}-\frac{pk}{m_0}} -\int_0^\Lambda \frac{dq}{2\pi^2} q^2   \frac{m_0}{2pq} {\rm ln}\frac{E-\frac{p^2+q^2}{2m_r}+\frac{pq}{m_0}}{E-\frac{p^2+q^2}{2m_r}-\frac{pq}{m_0}} t_2(E-\frac{q^2}{2m_{AD}}) t_3(q,k;E)
  \label{STM}
\end{equation}
\end{widetext}
Here, $\Lambda$ is the momentum cutoff, and $m_{AD}=m_b(m_0+m_b)/(m_0+2m_b)$ is the reduced mass for atom-dimer scattering. Eq.~\ref{STM} can be viewed as the corresponding dissipative STM equation in the presence of two-body losses~\citep{STM, Braaten}. Similar to $t_2$, the poles of $t_3$ on the complex energy plane denote three-body bound states.
In the following, we take the $^{133}$Cs-$^{133}$Cs-$^6$Li as a concrete example to calculate the whole energy spectra of trimers in the presence of two-body losses.

\textit{Results.} As discussed previously, the trimer energies are in general complex functions of inverse complex scattering length $a_c^{-1}$, which may be tuned across the upper half complex plane via the optical Feshbach resonance. It is thus natural to choose the argument and modulo of $a_c^{-1}$ as independent tuning parameters, as it gives a much more clear picture of the complex structures of the energy functions.

For example, if one checks the energy trajectories around the uniarity in Fig.~\ref{fig1}, one can see that the trajectories approximately lie on rays with angle $\pi-\phi$ for different $\phi=\text{arg}(a_c^{-1})$. This is because one may generalize Tan's adiabatic relation
\begin{equation}
   \frac{\partial E}{\partial a_c^{-1}} =-\frac{\hbar^2}{16\pi m_r} C
\end{equation}
to complex energies and scattering lengths~\cite{Contact}. Here, $C$ is the two-body contact, and $\hbar$ is the reduced Planck's constant, with being set to unity for simplicity. Around the unitarity $a_c^{-1}=0$, the trimer energies can be expressed as $ \Delta E \propto C(0)\Delta a_c^{-1}$, with $C(0)$ being the two-body contact at unitarity. This indicates the trimer energies must move along the directions with angles $\pi-\phi$ in the complex energy plane~\citep{SM}. On the other hand, if we tune $|a_c^{-1}|$ away from unitarity, there generally exist two scenarios, which will be discussed in detail as follows.

\begin{figure}[t] 
    \centering
    \includegraphics[width=0.42\textwidth]{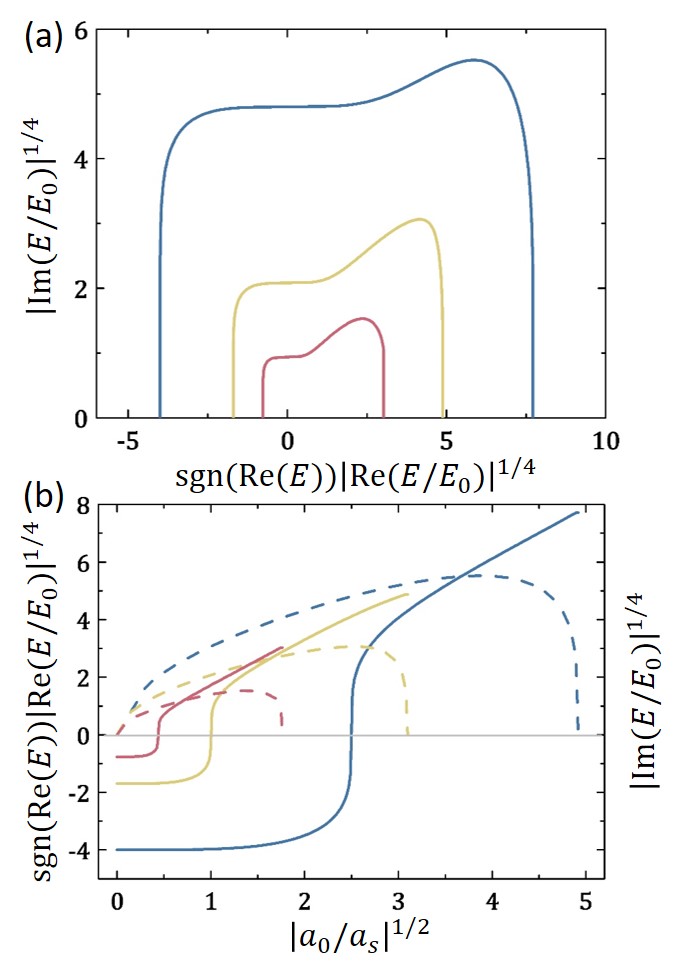}
    \caption{(a) The energy trajectories start from the three lowest Efimov trimers along $\phi =\pi/2$, with the first one (blue), the second one (yellow) and the third one (red). (b) The real (solid lines) and imaginary parts (dashed lines) of trimer energies plotted as functions of $|a_c^{-1}|$ ($\phi=\pi/2$ is fixed). $E_0$ and $a_0$ are the energy and length unit. The momentum cutoff is set to $\Lambda=20/a_0$ in this paper.}
     \label{fig2}
\end{figure}

We first take the argument $\phi=\pi/2$ as an example to investigate the first scenario in Fig.~\ref{fig1}, as well as the effect of two-body loss on the Efimov trimer. The three lowest branches of dissipative Efimov trimers are shown in Fig.~\ref{fig2}. When $|a_c^{-1}|=0$, there exist a series of conventional Efimov trimers with Im$(E)=0$. As $|a_c^{-1}|$ increases, the trimer energies become complex. Note that their imaginary parts are always negative in the presence of two-body losses. The inverse of imaginary part of $E$ gives a time scale which represents the lifetime of the trimer. While the real parts increase continuously, the imaginary parts exhibit a non-monotonic behavior, and vanish for large enough $|a_c^{-1}|$, where the trimers merge into three-body scattering states with positive energies. This is owing to the varying two-body loss rate $\gamma$, which can be expressed as 
\begin{equation}
   \gamma=\frac{1}{2\pi m_r}\frac{|a_c^{-1}|}{|a_c^{-1}|^2+(\frac{\Lambda}{\pi})^2}
\end{equation}
As one can see, it also displays a non-monotonic behavior as $|a_c^{-1}|$ increases.

Fig.~\ref{fig2}(a) also shows that different branches of complex Efimov trimers display an intriguing discrete scaling behavior. That is, if we scale both the real and imaginary part of the $n$-th branch by $e^{-2\pi/s_0}$, it matches the $n+1$-th energy branch. This can be viewed as a complex generalization of the celebrated Efimov's radial law~\cite{Efimov,Braaten}. To see this, we may apply the hyperspherical coordinate approach~\cite{Nielsen} for the three-body problem and consider the hyperspherical adiabatic potential $U(\rho)$ with $\rho$ being the hyperradius. Note that in the region $\Lambda^{-1}\ll \rho\ll |a_c|$, the system can be essentially viewed as at unitarity. As a result, we know that the different hyperspherical adiabatic channels are decoupled in this region and the lowest potential channel is $U(\rho)=-(s_0^2+1/4)/(2 \rho^2)$. It is then straightforward to show that the solutions to the three-body problem possess following radial law,
\begin{align}
a_c^{-1}\rightarrow e^{-\pi/s_0}a_c^{-1},\qquad E\rightarrow e^{-2\pi/s_0}E,
\label{scaling}
\end{align}
if one follows the conventional argument for the Efimov radial law~\cite{Efimov,Braaten}. And this law immediately leads to the discrete scaling behavior shown in Fig.~\ref{fig2}(a). Similar results has been obtained by calculations through Born-Oppenheimer approximation and adibiatic hyperspherical approximation in a recent work~\cite{Cui2}.

Another interesting feature is the existence of trimers with the real parts of their energies beyond the three-body scattering threshold ($\text{Re}(E)>0$). Note that the system do not support two-body bound states at $\phi=\pi/2$. The only scattering continuum is the three-body continuum, i.e. the positive real axis $\mathbb{R}^+$. Thus all the energy trajectories must end on the positive real axis, if assuming that the energy trajectories are continuous. This therefore allows trimers with the real parts of energies exceeding zero while with nonvanishing imaginary parts. Indeed, one can see from Fig.~\ref{fig1} and \ref{fig2}, when the real parts equal to zero, the trimers do not vanish at the three-body scattering threshold as in the conventional Efimov effect.
Moreover, the system can even support a series of trimers with energies that have positive real parts and very small imaginary parts. We note that these long-lived bound states are purely dissipation induced and the mechanism is completely different from other positive energy bound states such as bound state in the continuum~\cite{BOC}. We expect that these long-lived trimers can be detected in the experiments of ultracold atoms.

\begin{figure}[t] 
    \centering
    \includegraphics[width=0.48\textwidth]{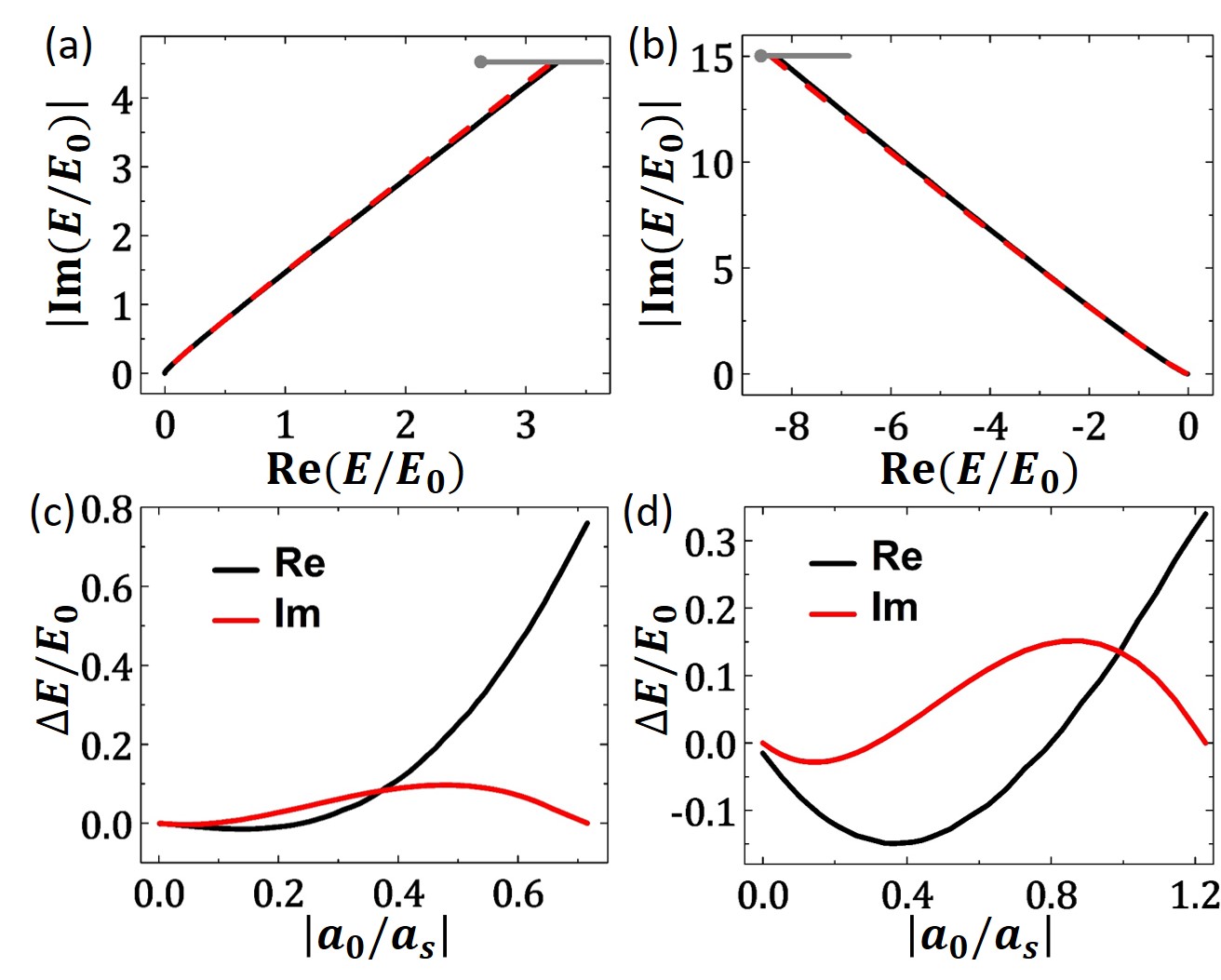}
    \caption{{\bf First row}: Energy trajectories in the complex plane. (a) $\phi=\pi/6$, the trajectories connected to the fifth (black) and the sixth (red dashed) real Efimov bound states. (b) $\phi=\pi/3$, the trajectories connected to the fourth (black) and the fifth (red dashed) real Efimov bound states.  Gray lines denote the atom-dimer continuum wheres the trimers disappear at.
    All red dashed lines are scaled by a factor of $e^{2\pi/s_0}\simeq23.7$ to help demonstrate the generalized radial law (Eq.~\ref{scaling})~\cite{footnote}.
    {\bf Second row}: $\Delta E=E-E_d$ as functions of inverse complex scattering length $|a_c^{-1}|$ with (c) $\phi=\pi/6$, and (d) $\phi=\pi/3$.  }
     \label{fig3}
\end{figure}

For $\phi < \pi/2$, the system supports a two-body heavy-light bound state with complex energy $E_d=-1/(2m_ra_c^2)$. And we find that the behavior of the energy trajectories changes qualitatively, even though different energy branches still show the generalized Efimov radial law (Eq.~\ref{scaling}) as demonstrated in Fig.~\ref{fig3} (a). 
Take $\phi=\pi/3$ and $\phi=\pi/6$ as examples, we see that the trimer energies exhibit the same feature once we subtract the dimer energy and plot $\Delta E=E-E_d$ as functions of $|a_c^{-1}|$. As shown in Fig.~\ref{fig3} (b), the real part of $\Delta E$ first decreases and then increases to a positive value, while the imaginary part displays an approximate S-shape, with two zero-value ending points. In contrast, the dissipative Efimov branches do not disappear at the atom-dimer scattering threshold, but at atom-dimer scattering states with finite scattering energies.

In particular, we find that $|\text{Im}(E)|<|\text{Im}(E_d)|$ for large $|a_c^{-1}|$, which means the trimer states have longer lifetimes than the dimers in this region. This effect is quite counter-intuitive as it implies that a two-body system can become more stable and long-lived by adding a third particle even in the presence of two-body losses. To understand this result, we adopt the Born-Openheimer approximation and calculate the light particle induced effective potential $V_{\text{eff}}(R)$ between the two heavy particles. Note that the imaginary part of $V_\text{eff}$ can be viewed as an effective two-body loss induced by the light particle. And indeed, we find that it is possible to have $|\text{Im}(V_\text{eff})|<|\text{Im}(E_d)|$ in some region, which suggests that the effective loss rate in an three-body system might be even smaller than the rate in a two-body system. We argue that this is a quantum mechanic effect which arises from the interference of the light particle wavefunctions~\cite{SM}. 

\begin{figure}[t] 
    \centering
    \includegraphics[width=0.42\textwidth]{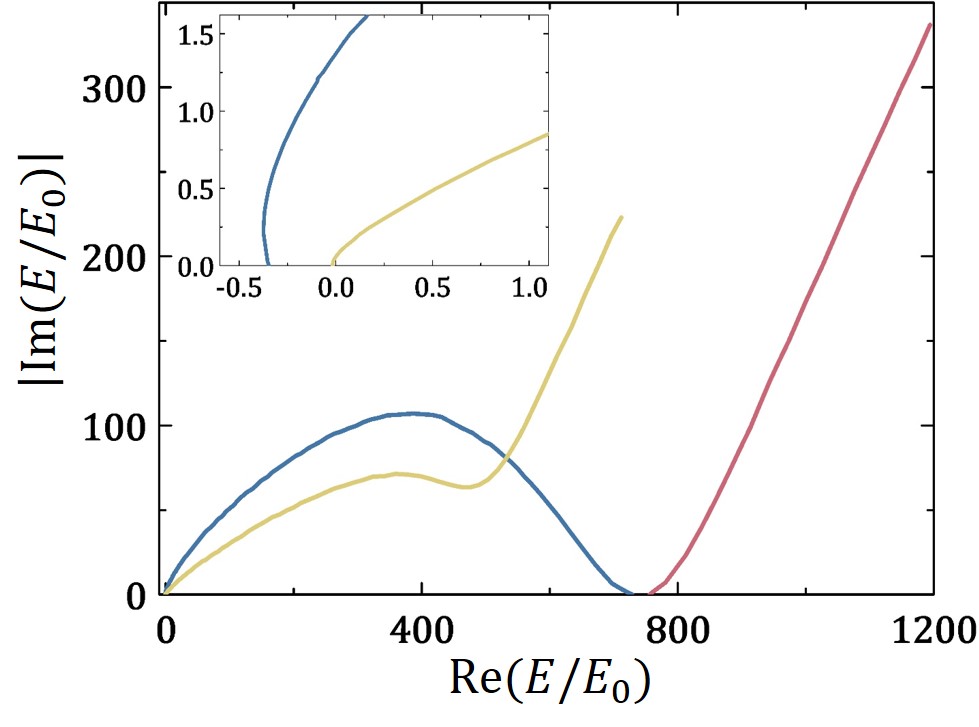}
    \caption{ Disconnected branches at $\phi\simeq0.423\pi$. The third lowest Efimov branch (blue) ends at Im$(E)=0$, while the fourth one (yellow) ends at an atom-dimer scattering state with Im$(E)\neq 0$. There exists another branch (red) near the third one, which starts at a three-body scattering state with Im$(E)=0$ and ends at an atom-dimer scattering state with Im$(E) \neq 0$. The inset is the zoom-in of the spectra around the origin, to demonstrate that the third and the fourth branches connect to the conventional Efimov trimers. }
     \label{fig4}
\end{figure}

\textit{Disconnected branches.} It is worth noting that the two scenarios discussed above only apply to low energy shallow bound states ($|E|\ll \Lambda^2/(2m_r)$). For deeper bound states, the system is more complex and non-universal as the effect of momentum cutoff becomes more and more important.
When the argument $\phi$ is slightly smaller than $\pi/2$, several deepest energy branches do not vanish into the atom-dimer scattering state continuously, but break down into two branches near the positive real energy axis, as displayed in Fig.~\ref{fig4}. The branch that connected to the regular dissipationless Efimov state ends at a three-body scattering state. While the other disconnected branch appears slightly above it along the real axis, which further extends to the complex plane and eventually disappears into an atom-dimer scattering state. In fact, for the $n$-th Efimov branch we find there exists a critical argument $\phi_{c}^{(n)}$. The two branches disconnect as the blue and red lines shown in Fig.~\ref{fig4} when $\phi>\phi_{c}^{(n)}$. While they connect to each other and form one branch when $\phi\leq\phi_c^{(n)}$ as displayed by the yellow line. We also find that the critical arguments $\phi_c^{(n)}$ increases while $n$ increases and it converges to $\pi/2$ for $n\rightarrow+\infty$. It was worth noting that similar critical argument phenomena also appear in the calculation of Born-Oppenheimer effective potential~\cite{SM}.

\textit{Conclusions.} 
We study the influence of short-range two-body losses on the three-body Efimov bound states. In contrast to the bound states in closed systems whose energies are always constrained on the real axis, the Efimov trimers can have complex energies in the presence of two-body losses. This fact leads to a series of distinct features with no counterparts in the conventional dissipationless Efimov physics. First, we find that the trajectories of Efimov trimers display a unique discrete scaling behavior in the complex energy plane, which is a complex analog of the celebrated Efimov radial law. Second, we show that the system can support Efimov bound states with positive energies and infinite long lifetimes even in the presence of two-body losses. Third, we show that in some region, the dissipative three-body system may have even longer lifetime than the two-body system. We argue that this counter-intuitive phenomenon is because of the interference effect of the light particle wavefunctions.

The work opens a new avenue toward the study on novel few-body physics in the complex energy plane.

\textit{Acknowledgment.} We thank Xiaoling Cui, Hui Zhai, Ran Qi, Ren Zhang, Lei Pan, Lihong Zhou for inspiring discussions. The project was supported by Fund of State Key Laboratory of IPOC (BUPT) No. 600119525 and 505019124 (MYS), NSFC under Grant No. 12004049 (MYS), 12004115 (ZYS),  Program of Shanghai Sailing Program Grant No. 20YF1411600 (ZYS).

\begin{widetext}

\section*{Supplementary material: Efimov physics in the complex plane}

\subsection*{Trimer energies near unitarity}
In Fig.~\ref{figS2}, we plot the Efimov energies for different $\phi=\text{arg}(a_c^{-1})$ near unitarity. Note that in this region we have
\begin{align}
  \Delta{E}\simeq-\frac{\hbar^2}{16\pi m_r}C(0) a_c^{-1},
\end{align}
where $\Delta E\equiv E-E(0)$ is the trimer energy measured from the Efimov state energy at unitarity. We thus conclude that $|\Delta E|\propto|a_c^{-1}|$ and $\text{arg}(\Delta E)\simeq\pi-\phi$. Both relations are demonstrated in Fig.~\ref{figS2}.

\begin{figure}[htp] 
    \centering
    \includegraphics[width=.9\textwidth]{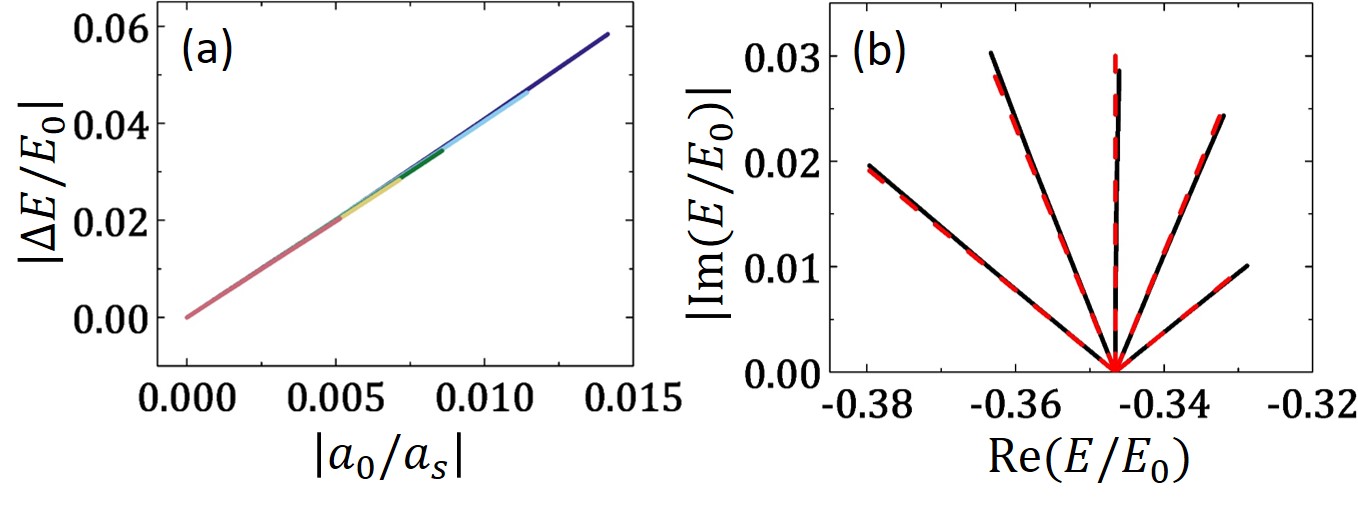}
    \caption{(a) $|\Delta E|\equiv |E-E(0)|$ as functions of $|a_c^{-1}|$, different colors correspond to different arguments of $a_c^{-1}$, including $\phi=\pi/6,\pi/3,\pi/2,2\pi/3,5\pi/6$ (the same as in Fig.~\ref{fig1}). One can see that all the lines collapse on a single ray indicating $|\Delta E|\propto|a_c^{-1}|$. (b) Energy trajectories near unitarity for different arguments $\phi=\pi/6,\pi/3,\pi/2,2\pi/3,5\pi/6$ (Black) and rays with corresponding direction angles $\pi-\phi$ (Red dashed). The plot demonstrates that $\text{arg}(\Delta E)\simeq\pi-\phi$.}
     \label{figS2}
\end{figure}

\subsection*{Born-Oppenheimer approximation for three-body problem with complex scattering length}

The Born-Oppenheimer potential $\epsilon(R)$ is obtained by calculating the eigenenergy of the light particle (with mass $m$) within the presence of two fixed heavy particles (with mass $M$) apart with distance $R$. The Schr\"{o}dinger equation for the light particle is given by
\begin{equation}
-\frac{\hbar^2}{2m}\nabla_\textbf{r}^2\psi(\textbf{r}) = \epsilon(R)\psi(\textbf{r})
\end{equation}

We assume the two heavy particles are located at $-\textbf{R}/2$ and $\textbf{R}/2$. Then we have following boundary conditions
\begin{equation}
\psi(\textbf{r}) \approx \frac{1}{\left|\textbf{r}\pm\frac{\textbf{R}}{2}\right|}-\frac{1}{a_c},\quad \textbf{r}\to\pm\frac{\textbf{R}}{2}
\end{equation}

Because of the parity symmetry of the problem, it is straightforward to show that the solution to the Schr\"{o}dinger equation is
\begin{equation}
\psi_\pm(\textbf{r}) = G_\epsilon(\textbf{r}-\frac{\textbf{R}}{2})+ G_\epsilon(\textbf{r}+\frac{\textbf{R}}{2})
\end{equation}
with 
\begin{equation}
G_\epsilon(\textbf{r}) = \frac{e^{-\kappa r}}{r}
\end{equation}
  being the Green's function. Here we assume $\kappa = \sqrt{-2m\epsilon/\hbar^2}$ with the branch cut of $\sqrt{.}$ defined on the negative real axis. Furthermore it is important that Re$(\kappa)\geq0$ such that $\psi$ can be normalized.
  
Substitute this into the boundary conditions, we immediately have
\begin{equation}
t- e^{-t} = \frac{R}{a_c}
 \label{BO}
\end{equation}
with $t = \kappa R$.
\begin{figure}[h] 
    \centering
    \includegraphics[width=0.7\textwidth]{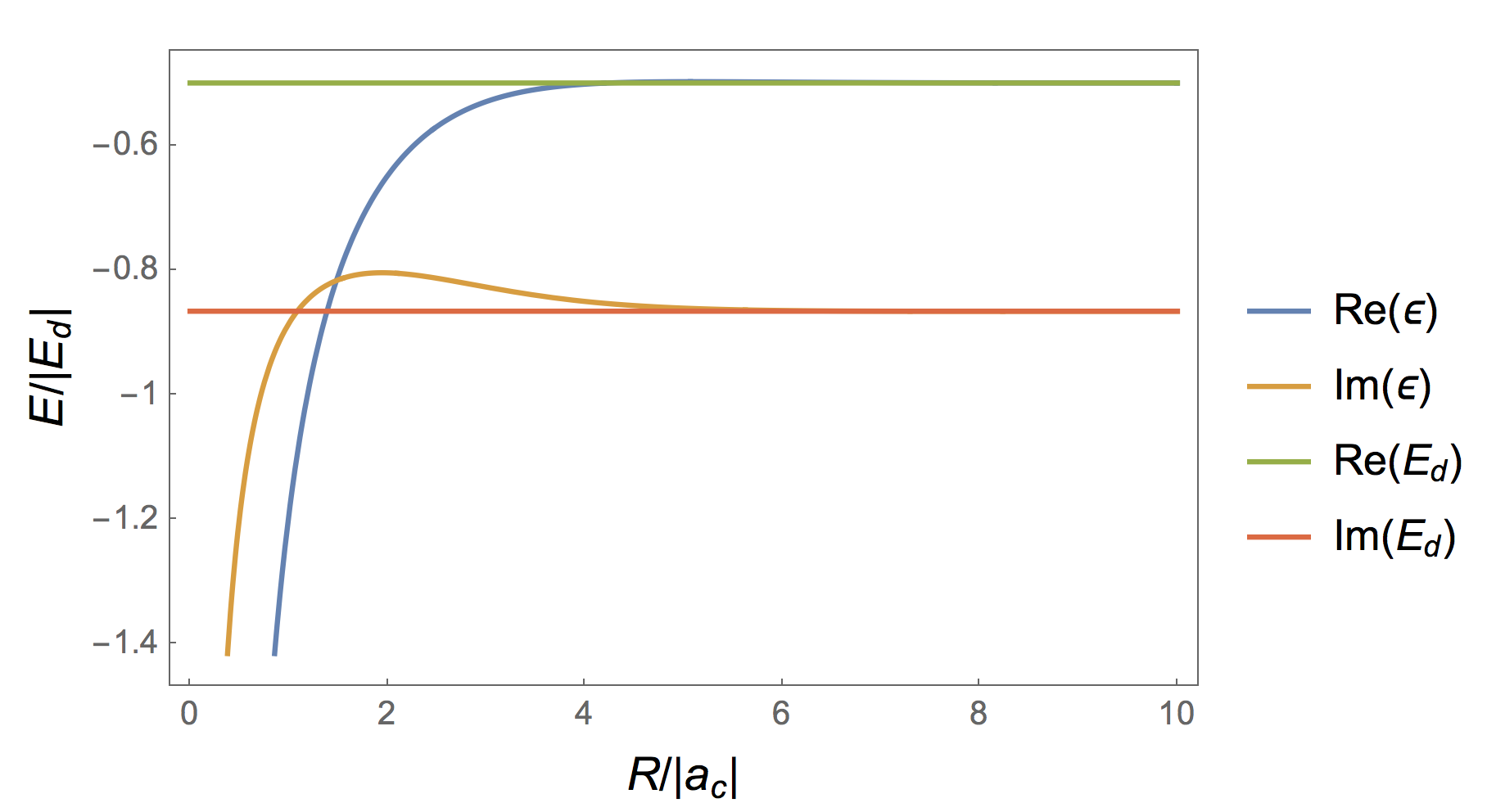}
    \caption{ The real and imaginary part of effective potential $\epsilon(R)$ and two-body bound state energy with $\left|a_c^{-1}\right| = 1$ and $\arg(a_c^{-1}) = \pi/6$. $V_r$ and $V_i$ tend to $E_r$ and $E_i$ at large distance but not always smaller than them like in closed system.  }
     \label{figS1}
\end{figure}

In Fig.~\ref{fig1}, we plot the real and imaginary part of effective potential $\epsilon(R)$ by numerically solving Eq.~\ref{BO}. As mentioned in the main text, we find that $|\text{Im}(\epsilon(\mathbf{R}))|<|\text{Im}(E_d)|$ in some region. In fact, one can show that $\text{Im}(\epsilon(R))\simeq-\text{Im}(E_d)+\frac{1}{m|a_c|R}e^{-R\cos\phi/|a_c|}\sin(\frac{R\sin\phi}{|a_c|}-\phi)$ for $R\gg|a_c|$, which is a function that oscillates around $\text{Im}(E_d)$. Clearly this oscillation arises from the interference between wave functions $G_\epsilon(\mathbf{r}-\frac{\mathbf{R}}{2})$ and $G_\epsilon(\mathbf{r}+\frac{\mathbf{R}}{2})$, which is purely quantum mechanic.

{\it Critical angles.} Numerically, we find that Eq.~\ref{BO} either has one or no root with $\text{Re}(\kappa)>0$, depends on the values of $R/a_c$. For $\phi=\text{arg}(a_c^{-1})<\phi_{c}^{(1)}\approx 1.26263$, Eq.~\eqref{BO} always has one solution for all $R>0$, which means that the light particle always induces an effective binding between heavy particles. For $\phi\in[\phi_{c}^{(1)},\phi_{c}^{(2)})\approx[1.26263,1.46509)$, Eq.~\ref{BO} has no solution in a bounded interval around $R/|a_c|=3$. In this region, the heavy particles tend to bind when they are far apart, but this binding effect induced by the light particle vanishes around $R/|a_c|=3$. For $\phi>\phi_{c}^{(2)}$, a second interval near $R/|a_c|=9$ appears which represents that the binding effect vanishes in it. Keep increasing $\phi$, we find that at $\phi=\pi/2$, there appear an infinite number of intervals where Eq.~\ref{BO} has no solutions. This indicating that $\lim_{n\rightarrow+\infty}\phi_c^{(n)}=\pi/2$ and the Born-Oppenheimer approximations fails even for very large $R$.

Based on these analysis, it is thus natural to expect that for $\phi\in(\phi_c^{(1)},\pi/2)$, some ``deep'' Efimov states may end on the three-body continuum while all ``shallow'' Efimov states end on the atom-dimer continuum. This is also in consistent with our numerical results.


\end{widetext}


\begin{thebibliography}{99}
\bibitem{footnote0}
This is nontrivial as one might have $\text{Re}(a_c)=0$ but the system is still at two-body resonance due to two-body loss.
\bibitem{footnote0.5}
We note that similar results have been obtained by authors of ref.~\cite{Iskin,Cui2}, in which an extra restriction $\text{Re}(E_d)<0$ is posed. While we find the restriction unnecessary. Indeed, one may find an eigen state with square integrable wave function $\psi(r)=\frac{e^{-r/a_c}}{r}$ for all the complex scattering length with $\text{Re}(a_c)>0$, which inevitably results in bound states with energies $\text{Re}(E_d)\geq0$.
\bibitem{footnote}
Note that the momentum cutoff can significantly revise the energy of the several lowest branches, similar as in the conventional Efimov effect, which lead to the breakdown of the scaling law quantitatively.
\end{thebibliography}
\end{document}